\documentclass[10pt,twocolumn]{article}

\usepackage[left=48pt, right=42pt,top=46pt,bottom=60pt,headheight=15pt,headsep=10pt,letterpaper,twoside]{geometry}
\usepackage[activate={true,nocompatibility},final,tracking=true,kerning=true,spacing=,factor=1100,stretch=10,shrink=10]{microtype}  
\usepackage[english]{babel}  
\usepackage{booktabs}  
\usepackage{makecell}  
\usepackage[font=small,labelfont=bf]{caption}  
\usepackage{float}  
\usepackage{graphicx,xcolor,tabularx}  
\graphicspath{{figures_main/}}  
\usepackage[section]{placeins}  
\newcommand{\xsection}[1]{\medskip\noindent{\large{\textbf{#1}}}\newline}  
\newcommand{\xsubsection}[1]{\noindent{\textbf{#1}}}
\usepackage{lineno}  
\usepackage[colorlinks,linkcolor=blue,citecolor=blue]{hyperref}  
\usepackage{paralist}  
\usepackage[superscript,compress,nomove]{cite}  
\newcommand{\refpath}{../../../MyLibrary}
\usepackage{authblk}  

\usepackage{amsmath,amsfonts,amsthm,amssymb,mathrsfs}  
\usepackage{siunitx}  
\usepackage{algorithm}  
\usepackage{algpseudocode}  
\DeclareMathOperator*{\argmin}{arg\,min}  
\renewcommand{\vec}[1]{\ensuremath{\mathbf{#1}}}  
\newcommand{\abs}[1]{\ensuremath{\left\lvert #1 \right\rvert}}
\newcommand*{\transp}{^{\intercal}}

\newcommand*{\vpsf}{PSF\textsubscript{vec}}
\newcommand*{\ipsf}{PSF}

\newcommand{\insitu}{\textit{in situ}}
\newcommand{\Insitu}{\textit{In situ}}

\newcommand{\supp}[1]{\textcolor{blue}{#1}}

\title{\Insitu{} fully vectorial tomography and pupil function retrieval \\of tightly focused fields}

\author[1,2]{Xin Liu}
\author[1]{Shijie Tu}
\author[1]{Yiwen Hu}
\author[2]{Yifan Peng}
\author[1]{Yubing Han}
\author[1]{\\Cuifang Kuang}
\author[1,*]{Xu Liu}
\author[1,*]{Xiang Hao}

\affil[1]{College of Optical Science and Engineering, Zhejiang University, Hangzhou 310027, China}
\affil[2]{Department of Electrical and Electronic Engineering, The University of Hong Kong, Hong Kong SAR, China}
\affil[*]{Corresponding author: haox@zju.edu.cn; liuxu@zju.edu.cn;}

\date{}
\begin{document}
\maketitle

\xsubsection{Abstract:}
Tightly focused optical fields are essential in nano-optics, but their applications have been limited by the challenges of accurate yet efficient characterization.
In this article, we develop an \insitu{} method for reconstructing the fully vectorial information of tightly focused fields in three-dimensional (3D) space, while simultaneously retrieving the pupil functions.
Our approach encodes these fields using phase-modulated focusing and polarization-split detection, followed by decoding through an algorithm based on least-sampling matrix-based Fourier transform and analytically derived gradient.
We further employ a focus scanning strategy. When combined with our decoding algorithm, this strategy mitigates the imperfections in the detection path.
This approach requires only 10 frames of 2D measurements to realize approximate \qty{90}{\percent} accuracy in tomography and pupil function retrieval within \qty{10}{\second}.
Thus, it serves as a robust and convenient tool for the precise characterization and optimization of light at the nanoscale.
We apply this technique to fully vectorial field manipulation, adaptive-optics-assisted nanoscopy, and addressing mixed-state problems.

\xsection{Introduction}
Paraxial beams behave like transverse waves with two-dimensional (2D) polarization orthogonal to the propagation axis. However, when tightly focused by high numerical-aperture (NA) objectives, they always exhibit an evident polarization component along the propagation direction~\cite{Quabis2000FocusingLightTighter, Wang2008CreationNeedleLongitudinally, Zhan2009CylindricalVectorBeams}, resulting in a 3D polarization structure. This 3D polarization plays a crucial role in studying light behavior at nanometer scales, such as spin-to-orbital angular momentum conversion~\cite{Zhao2007SpintoorbitalAngularMomentum, Man2020DualCoaxialLongitudinal} and topological photonics~\cite{Bauer2015ObservationOpticalPolarization, Maucher2018CreatingComplexOptical, Zeng2024TightlyFocusedOptical}. It also underpins various applications like optical tweezers~\cite{Beguin2020ReducedVolumeReflection, Wu2024ControllableMicroparticleSpinning}, laser manufacturing~\cite{Li2022RealisingHighAspect}, spectroscopy~\cite{Grosche2020PolarizationbasedExcitationTailoring}, and optical nanoscopy~\cite{Balzarotti2017NanometerResolutionImaging, Hao2021ThreedimensionalAdaptiveOptical}. To fully harness its potential, it is vital to accurately and comprehensively characterize these fields, i.e., to capture the fully vectorial information---both the amplitude and phase of all three polarization components. Meanwhile, knowing the pupil function defined on the objective's back focal plane (BFP) is critical for manipulating and optimizing these fields in experiments. Unfortunately, achieving both aims has been challenging due to the absence of appropriate tools.

Recent advancements have attempted to measure the tightly focused fields through direct magnification~\cite{Herrera2023MeasurementStructuredTightly, Quinto-Su2023InterferometricMeasurementArbitrary, Maluenda2021ExperimentalEstimationLongitudinal, Martinez-Herrero2023LocalCharacterizationPolarization}. It is also feasible to use a probe to detect the near field~\cite{Grosjean2010FullVectorialImaging} or map it to the far field~\cite{Bauer2014NanointerferometricAmplitudePhase, Yang2023MieScatteringNanointerferometry}. However, these methods face limitations due to imperfections in the detection path and require complicated setups. In addition, the far-field mapping needs to capture thousands of images for reconstruction, which are especially time-consuming~\cite{Neugebauer2014GeometricSpinHall, Yang2023MieScatteringNanointerferometry}. These techniques are further constrained as they are only available in a 2D plane near the focus, with the 3D reconstruction unrealized. Moreover, the \insitu{} characterization of the associated pupil functions remains unexplored.

To solve these problems, we introduce an approach for reconstructing both the fully vectorial information of tightly focused fields in 3D space and the associated pupil functions using optical encoding and algorithmic decoding. Specifically, the vectorial information is encoded in 2D point spread functions (\ipsf{}s) via active phase modulation and polarization-split detection. The information is subsequently decoded using Fourier-transform-based analytical gradient descent. By employing focus scanning to detect the \ipsf{}s and leveraging our decoding algorithm, we mitigate the imperfections in the detection path, enabling \insitu{} tomography and pupil function retrieval.

\begin{figure*}[!t]
    \centering
    \includegraphics{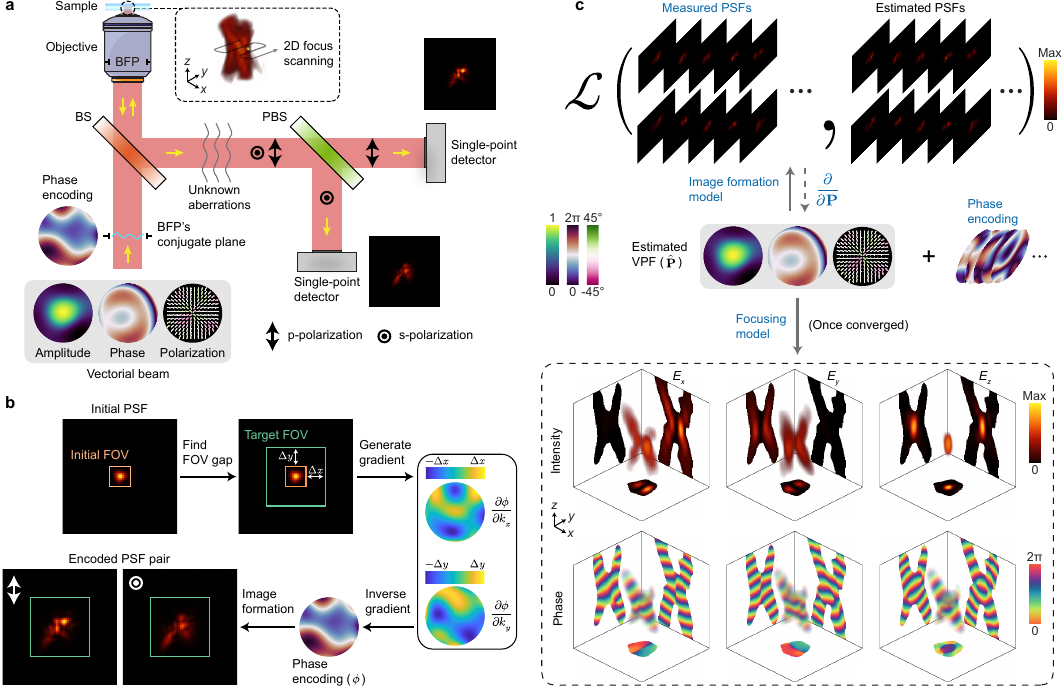}
    \caption{\textbf{Concept of \insitu{} fully vectorial tomography and pupil function retrieval.}
        \textbf{a}, Schematic of optical encoding. The phase encoding is applied on the conjugate plane of the objective’s BFP. Yellow arrows indicate the direction of light propagation. Abbreviations: BFP, back focal plane; BS, beamsplitter; PBS, polarizing beamsplitter.
        \textbf{b}, Phase encoding design. We first define the boundary (orange box) of a pre-captured \ipsf{} and the target field of view (FOV) (green box) that encloses the encoded \ipsf{}. The gap between the two boundaries correlates with the phase gradient's magnitude of the VPF. We randomly generate the phase gradient within this magnitude constraint, and then derive the phase by computing its inverse gradient. Applying this phase encoding yields the encoded two-channel \ipsf{}s within the target FOV. Further details are available in the Methods section.
        \textbf{c}, Schematic of algorithmic decoding. Blue text represents known parameters or operators. Gray solid and dashed arrows indicate the forward process and the update of the VPF, respectively. The polarization's colormap illustrates the ellipticity.
        }
    \label{fig_principles}
\end{figure*}

\xsection{Principle}
Our method reconstructs the tightly focused field (\vpsf{}) in 3D space and the corresponding vectorial pupil function (VPF) from several measured 2D \ipsf{}s. The whole process is to solve a phase-retrieval problem~\cite{Fienup1982PhaseRetrievalAlgorithms, Kromann2012QuantitativePupilAnalysis, HieuThao2020PhaseRetrievalBased, Vishniakou2023DifferentiableOptimizationDebyeWolf, Gutierrez-Cuevas2024VectorialPhaseRetrieval}. However, phase retrieval is ill-posed, because different complex fields can produce identical measurements. The challenge is more profound in \vpsf{} reconstruction due to the inclusion of polarization, especially since the three polarization components are interconnected by the VPF. To eliminate this ambiguity, we implement a joint focusing and detection encoding strategy~(Fig.~\ref{fig_principles}a). In detail, we apply phase diversity as the focusing encoding~\cite{Gonsalves2018PhaseDiversityMath, Wu2020WISHEDWavefrontImaging}, where extra phase modulation with prior knowledge is introduced to VPF, thereby redistributing the \vpsf{}. Unlike phase diversity realized by defocus or vanilla random modulation~\cite{Gonsalves2018PhaseDiversityMath, Wu2019WISHWavefrontImaging, Gutierrez-Cuevas2024VectorialPhaseRetrieval}, we elaborately design these phases to balance diversity with signal-to-noise ratio (SNR) (Fig.~\ref{fig_principles}b). This balance is important because complicated VPFs compromise the SNR by over-distorting the PSF, whereas simple ones can not provide enough diversity to support the reconstruction. On the detection side, the \ipsf{}s are acquired by scanning a dipole-like probe, e.g., a gold nanosphere, and collecting the backscattered signal using single-point detectors. While phase diversity typically ensures unique solutions in scalar case~\cite{Fannjiang2012AbsoluteUniquenessPhase, Wu2020WISHEDWavefrontImaging}, the diversity can be unnoticeable when incorporating polarization, rendering the solution ambiguous again under limited SNR conditions. To mitigate this, we separately detect the p- and s-polarization components (\supp{Supplementary Section~1}). The polarization separation helps decouple the contributions of each 3D polarization component. To support subsequent algorithmic decoding, we establish a comprehensive image formation model for the entire process (\supp{Supplementary Section~2}).

Given the known phase encoding and the measured \ipsf{}s from two polarization channels, we algorithmically retrieve the VPF by solving the following inverse problem:
\begin{equation}
    \argmin_{\hat{\vec{P}}} \mathcal{L} \left\{\mathbb{I} \left(\vec{P}, \Psi \right), \widetilde{\mathbb{I}} \right\},
    \label{eq_principles_rec}
\end{equation}
where \(\Psi\) represents the known phases for encoding, while \(\mathbb{I}\) and \(\widetilde{\mathbb{I}}\) are the estimated and measured \ipsf{}s, respectively. The error function \(\mathcal{L}\) quantifies the difference between the estimated and measured \ipsf{}s, with \(\hat{\vec{P}}\) denoting the estimated VPF. To solve Eq.~(\ref{eq_principles_rec}), we develop an algorithm leveraging Fourier transform and analytically derived gradients (\supp{Supplementary Sections~3.1--3.2}). Once the VPF is obtained, the \vpsf{} in 3D space can be computed according to the focusing model~(Fig.~\ref{fig_principles}c).

Our method provides three major advantages over previous works~\cite{Herrera2023MeasurementStructuredTightly, Quinto-Su2023InterferometricMeasurementArbitrary, Maluenda2021ExperimentalEstimationLongitudinal, Martinez-Herrero2023LocalCharacterizationPolarization, Grosjean2010FullVectorialImaging, Bauer2014NanointerferometricAmplitudePhase, Yang2023MieScatteringNanointerferometry}:
\begin{compactitem}
    \item \textbf{Fully vectorial tomography and pupil function retrieval:} The joint focusing and detection encoding strategy enables unambiguous retrieval of the VPF. With the reconstructed VPF, we achieve the fully vectorial tomography of the \vpsf{} following the focusing model.
    \item \textbf{\Insitu{} characterization:} The focus scanning strategy directly probes the \vpsf{} near the focus without re-imaging. The single-point detector enables us to record the total energy collected by the objective rather than pixelated images for each scanning point. Therefore, it minimizes the impact of imperfections in the detection path, including amplitude, phase, and polarization aberrations~\cite{Chipman2018PolarizedLightOptical}. Besides the benefits of the hardware, our decoding algorithm eliminates imperfections by incorporating them into optimization, further enhancing the method's \insitu{} capabilities.
    \item \textbf{High efficiency:} Also benefiting from the optical encoding, only several 2D \ipsf{} acquisitions are required for accurate reconstruction. Moreover, during algorithmic decoding, we determine the least samples of the VPF and \vpsf{}, and use matrix-based Fourier transform~\cite{Liu2023FastGenerationArbitrary, Wei2023ModelingOffaxisDiffraction} to ensure accurate results with minimal computational effort.
\end{compactitem}
See \supp{Supplementary Sections~3.3--3.5} for further details.

\xsection{Results}
\xsubsection{Simulations.}
To quantify the performance of our method, we first define the reconstruction accuracy based on normalized root-mean-square-error (\supp{Supplementary Section~4.1}). Following this definition, we began with validating our method through simulations. Various tightly focused beams were successfully reconstructed at approximate \qty{90}{\percent} accuracy without failure case, including those highly ambiguous ones once our encoding strategy is not applied. Moreover, we also achieved around \qty{80}{\percent} accuracy even when the background noise is comparable to the signal. Remarkably, our decoding algorithm allows completing the reconstruction within \qty{10}{\second} on a standard commercial computer. Both in simulations and subsequent experiments, we employed 10 phases for encoding, determined based on the available SNR in our experiments (\supp{Supplementary Section~4.3}). Additionally, we determined an average peak intensity with 300 photons for the measured \ipsf{}s (120-nm pixel size) suffices for the reconstruction. Further details, including the accuracy definition, image processing, the influence of the number of phase encoding and the SNR, and the reconstructed results, are available in \supp{Supplementary Section~4}. This supplementary section also details the comparison between our method and earlier studies.

\medskip\xsubsection{Experiments.}
Following the simulations are the experimental validations. As direct experimental ground truth is unattainable, we alternatively confirmed the accuracy of the reconstructed results in two ways. First, we compared the amplitude and polarization of the reconstructed VPFs with those measured by a Stokes camera conjugate to the objective's BFP (\supp{Supplementary Section~5.1}). Second, we used the reconstructed \vpsf{} to predict the two-channel \ipsf{}s at different depths without phase encoding, and then compared them with the measured counterparts that did not contribute to the reconstruction. Notably, the VPFs measured by the Stokes camera only serve as a rough reference, as it is affected by the optical elements before the focus. In contrast, our reconstructed VPFs represent the ``effective'' counterparts that have included these effects. Moreover, compared to VPFs measured by the Stokes camera, the measured two-channel \ipsf{}s without phase encoding serve as a precise reference since the detection path remains the same as that for measuring phase-encoded \ipsf{}s.

\begin{figure*}[!t]%
    \centering
    \includegraphics{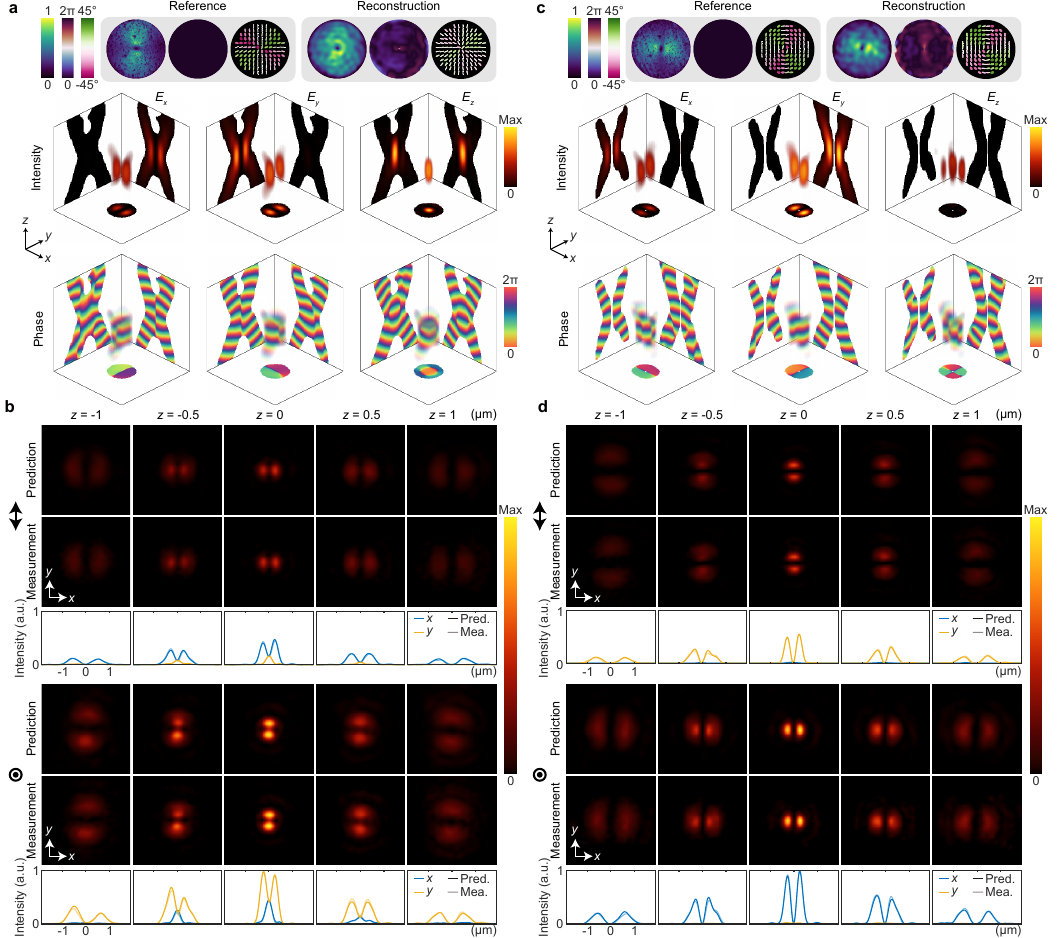}
    \caption{\textbf{Experimental results.}
        \textbf{a}, The reference VPF (top left), the reconstructed ones (top right), and the reconstructed \vpsf{} in 3D space with 2D projections (middle and bottom rows.) for the radially polarized beam. For visualization purpose, only regions with significant intensity are shown in the tomography, with volume size \(\left(\qty{3.84}{\um}\right)^3\).
        \textbf{b}, The predicted and measured two-channel \ipsf{}s without phase encoding at different depths for the radially polarized beam. The blue and yellow lines indicate intensity profiles along the \(x\) and \(y\) directions, respectively. Opaque lines represent the predicted \ipsf{}s, while translucent lines represent the measured ones. The measured \ipsf{}s are processed by low-pass filtering with the objective's bandwidth since the raw data suffers from Poisson noise (\supp{Supplementary Section~5.4}).
        \textbf{c} and \textbf{d} are for the azimuthally polarized beam.
        }
    \label{fig_psf_reconstruction}
\end{figure*}

In our experiments, we utilized a vortex half-wave plate (VHWP) to generate two cylindrical vector beams with radial and azimuthal polarization, respectively. These beams are ideal for validating 3D polarization reconstruction because the longitudinal component in their \vpsf{} is either dominant (for radial polarization) or negligible (for azimuthal polarization)~\cite{Youngworth2000FocusingHighNumerical}. Initially, we verified our method and corrected phase aberrations in the focusing path with linearly polarized light. The linear polarization (either p- or s-polarization) is less affected by polarization aberrations from optical elements~\cite{Chipman2018PolarizedLightOptical}, making the reference VPF more reliable (\supp{Supplementary Section~5.2}). Following this initial validation and aberration correction, we reconstructed the \vpsf{} and VPFs of the tightly focused cylindrical vector beams. Figures~\ref{fig_psf_reconstruction}a--d present the results. In detail, the reconstructed VPFs generally match the reference ones~(Figs.~\ref{fig_psf_reconstruction}a,c), including polarization singularities. Nevertheless, there are some noticeable discrepancies. Specifically, the amplitudes exhibit low-pass filtering due to the FOV limitation of the \ipsf{}s. The phases show some variations that ideally should be zero. These discrepancies are mainly attributed to the noise in measured \ipsf{}s and VPFs' polarization singularities, with additional phase aberrations potentially introduced by the VHWP after initial aberration correction. Other deviations in the VPFs are dominated by optical elements prior to the focus.

Despite discrepancies in the VPFs, the key features of the \vpsf{} for cylindrical vector beams are well preserved. For the radially polarized beam, the \vpsf{} exhibits nearly orthogonal shapes of \(E_x\) and \(E_y\) with a linear dark region and a phase step. Additionally, the longitudinal component \(E_z\) is stronger than the lateral components and exhibits a circularly symmetric intensity and phase around the optical axis. In contrast, although the tightly focused azimuthally polarized beam also shows orthogonal shapes of the lateral components, the longitudinal component is much weaker (ideally zero). In both cases, \(E_y\) is stronger than \(E_x\), which should ideally be equal. This is due to greater attenuation of the \(x\)-component of the VPFs' polarization compared to the \(y\)-component, as evidenced by the circular asymmetry of the VPFs' amplitude. Furthermore, the estimated \ipsf{}s with phase encoding align well with the measurements (\supp{Supplementary Section~5.3}), confirming the efficacy of our image formation model and decoding algorithm. Next, we used the reconstructed \vpsf{} to predict the measured two-channel \ipsf{}s without phase encoding, which, as mentioned before, serves as the precise ruler. Notably, in both cases, the predicted \ipsf{}s perfectly align with the measured ones at various depths (Figs.~\ref{fig_psf_reconstruction}b,d), reaching around \qty{90}{\percent} accuracy which is consistent with the simulations. Our prediction is further supported by reconstructing a tightly focused double helix beam, whose \ipsf{} has two rotating lobes, and the angle of rotation depends on the axial position (\supp{Supplementary Section~5.5}).

\begin{figure*}[!t]%
    \centering
    \includegraphics{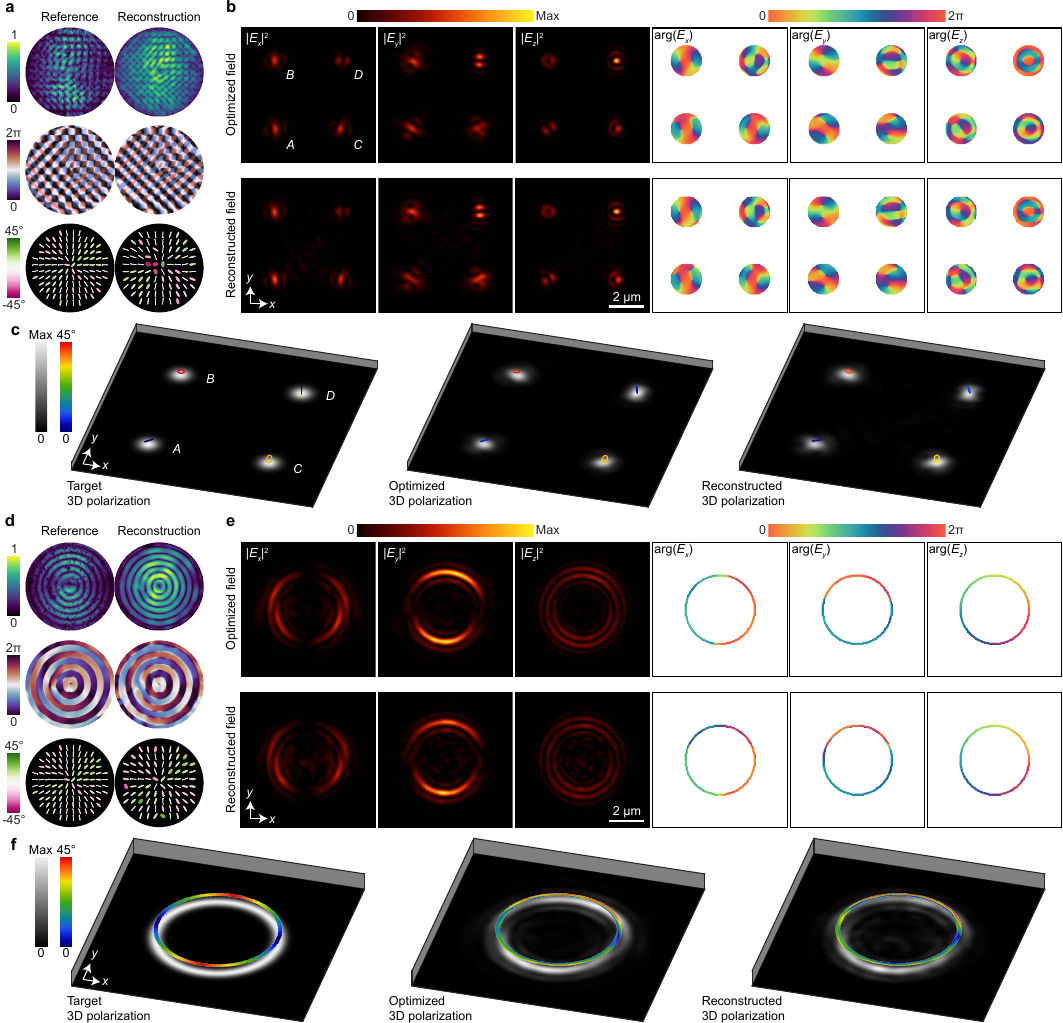}
    \caption{\textbf{Created novel complicated 3D polarization \insitu{}.}
        \textbf{a}, The reference VPF (left column), and the reconstructed one (right column). In the reference VPF, the middle row displays the phase optimized for the desired \vpsf{}. The top row illustrates the amplitude captured by the Stokes camera when the optimized phase is loaded onto a spatial light modulator (SLM). The bottom row shows the polarization constraint applied during phase optimization.
        \textbf{b}, The optimized \vpsf{} (top row), and the reconstructed counterpart (bottom row). Only regions with significant intensity are shown in the phase visualization. The electric fields of the target four foci are \(\vec{E}_A = \begin{bmatrix} 1/\sqrt{2}, 1/\sqrt{2}, 0\end{bmatrix}\transp\), \(\vec{E}_B = \begin{bmatrix} 1/\sqrt{2}, e^{i\pi/2}/\sqrt{2}, 0\end{bmatrix}\transp\), \(\vec{E}_C = \begin{bmatrix} 1/\sqrt{3}, e^{-i\pi/2}/\sqrt{3}, e^{-i\pi/2}/\sqrt{3}\end{bmatrix}\transp\), and \(\vec{E}_D = \begin{bmatrix} 0, 0, 1\end{bmatrix}\transp\), respectively, where \(\transp\) denotes matrix transpose.
        \textbf{c}, The polarization and intensity of the target (left column), the optimized (middle column), and the reconstructed (right column) \vpsf{}. The gray and rainbow colormaps indicate the intensity and the 3D polarization ellipticity, respectively. The normal directions of 3D polarization ellipses are omitted, as \(E_z\) may dominate.
        \textbf{d} -- \textbf{f} are for continuous \vpsf{}. The electric field of the target continuous field is \(\vec{E}(\vec{r}) = \begin{bmatrix} \cos{\varphi(\vec{r})}, \sin{\varphi(\vec{r})}, e^{i\varphi(\vec{r})}\end{bmatrix}\transp\), where \(\vec{r}\) and \(\varphi\) represent the position vector and the azimuthal angle on the focal plane, respectively.
        }
    \label{fig_psf_manipulation}
\end{figure*}

In these experimental validations, we have incorporated the imperfections of the detection path into the decoding process. The efficacy has been indicated by the reconstruction accuracy. To further illustrate the robustness of our detection strategy in the presence of aberrations, we introduced artificial aberrations by inserting a diffuser into the optical setup (\supp{Supplementary Section~5.6}). Specifically, when the diffuser was placed in the focusing path, it dramatically distorted the detected \ipsf{}s due to induced aberrations. In contrast, when the diffuser was in the detection path, the shape of detected \ipsf{}s remained stable despite a decrease in signal strength. This indicates that our detection strategy is robust against aberrations, which is crucial for \insitu{} characterization.

\begin{figure*}[!t]%
    \centering
    \includegraphics{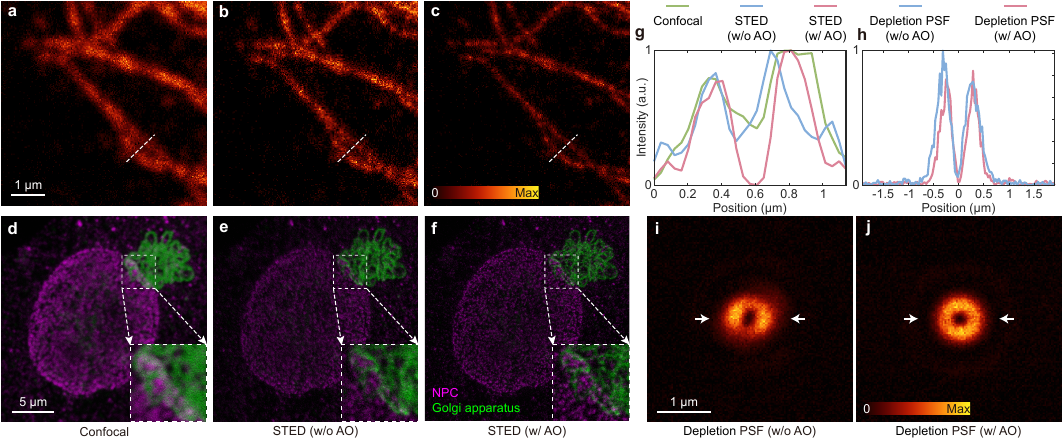}
    \caption{\textbf{STED imaging with \insitu{} aberration correction.}
        \textbf{a-c}, From left to right, these images show microtubules imaged using confocal microscopy, and STED nanoscopy without and with AO, respectively. Herein, the confocal image only serves as a rough reference to verify the STED images.
        \textbf{d-f}, Two-color imaging results of the NPC and Golgi apparatus.
        \textbf{g}, Intensity profiles along the white dashed lines in \textbf{a-c}, with slight smoothing applied for visualization.
        \textbf{h}, Intensity profiles along the arrows shown in \textbf{i} and \textbf{j}.
        \textbf{i} and \textbf{j} are the measured \ipsf{}s of the depletion beam with and without AO, respectively.
        }
    \label{fig_ao_sted}
\end{figure*}

\medskip\xsubsection{Application in field manipulation.}
Here we demonstrate that our method enables fully vectorial manipulation of the \vpsf{} \insitu{} using a gradient-based algorithm (\supp{Supplementary Sections~6.1--6.2}). In previous works, the amplitude and phase of the \vpsf{} can not be flexibly controlled~\cite{Urbach2008FieldFocusMaximum, You2015IterativePhaseretrievalMethod, Ren2020ThreedimensionalVectorialHolography, Abouraddy2006ThreedimensionalPolarizationControl, Chen2018VectorialOpticalFields, Liu2021GenerationArbitraryLongitudinal}. In addition, the \insitu{} manipulation remains missing. In contrast to these methods, we generate the desired \vpsf{} \insitu{} given the experimentally retrieved VPF. For this demonstration, we optimize the VPF's phase with the constraints of previously retrieved amplitude and polarization in Fig.~\ref{fig_psf_reconstruction}. We adopted radial polarization because it offers sufficient \(E_z\) strength, which facilitates the 3D polarization manipulation.

We first generated an optical focus array with predefined 3D polarization states (Figs.~\ref{fig_psf_manipulation}a--c), where the optimized \vpsf{} reaches \qty{76}{\percent} accuracy. Besides, the reconstructed VPF and \vpsf{} further verify the optimized counterparts in experimental realization. The dark areas in the VPF's amplitude are caused by phase singularities. Discrepancies in the reconstructed \vpsf{} are attributed to the amplitude and polarization constraints of the VPF during optimization since they vary with the phase in experiments when involving high-frequency features.

To demonstrate our method on a more challenging target, we applied our method to continuous fields (Figs.~\ref{fig_psf_manipulation}d--f). Compared with discrete foci, continuous fields may possess more rapid spatial variations (\supp{Supplementary Section~6.3}). Although the optimized \vpsf{} shows a ring-shaped intensity similar to the target, we are aware of a noticeable deviation in the 3D polarization (\qty{21}{\percent} accuracy). This phenomenon stems from phase-only modulation of the VPF. Better accuracy is possible if the modulation also includes amplitude and polarization. Nonetheless, the reconstructed VPF and \vpsf{} experimentally verify the optimized ones. Notably, the optimization for field manipulation is faster (within \qty{1}{\second}) compared to reconstruction (within \qty{10}{\second}), due to a more constrained optimization target.

\begin{figure*}[!t]%
    \centering
    \includegraphics{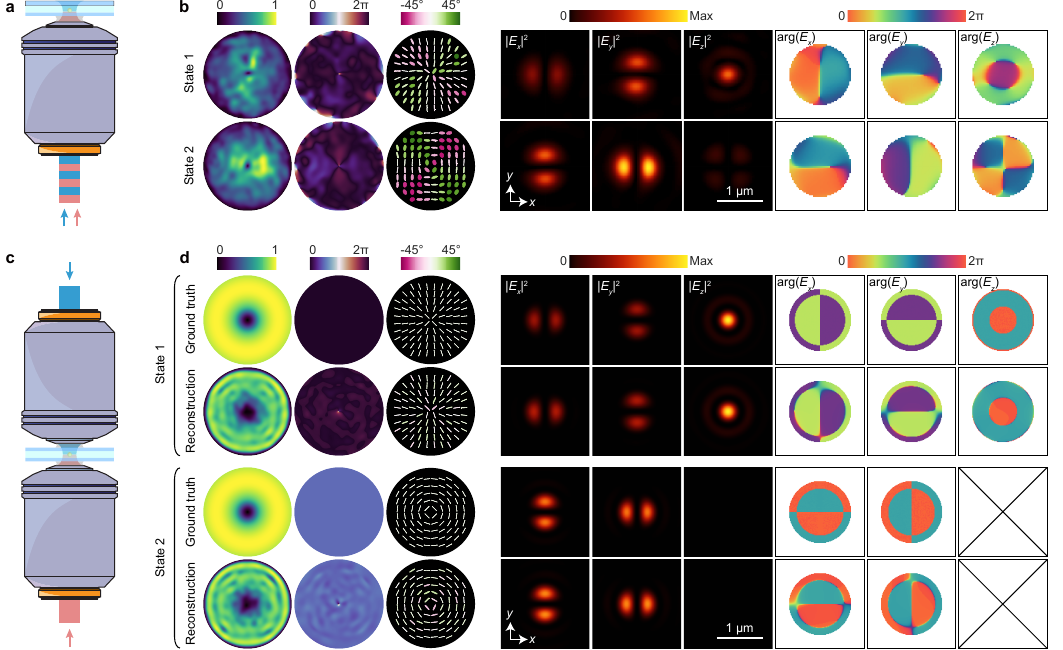}
    \caption{\textbf{Demixed results of two states.}
        \textbf{a}, Schematic illustration of incoherent superposition. The red and blue colors denote different beams.
        \textbf{b}, Experimental results for incoherent demixing. From left to right, the 1\textsuperscript{st}--3\textsuperscript{rd} columns denote the retrieved VPF; the 4\textsuperscript{th}--6\textsuperscript{th} and 7\textsuperscript{th}--9\textsuperscript{th} columns denote the intensity and phase of the reconstructed \vpsf{}, respectively. The phase encoding of the two states is achieved by simultaneously modulating different regions on the SLM to impose two distinct phases~\cite{Tu2022AccurateBackgroundReduction}.
        \textbf{c}, Schematic illustration of 4Pi focusing. In this configuration, two counter-propagating beams interfere near the shared focus of two opposing objectives. The resulting coherent mixed state consists of radially and azimuthally polarized beams with a phase difference.
        \textbf{d}, Simulated results for coherent demixing. From left to right, the 1\textsuperscript{st}--3\textsuperscript{rd} columns denote the VPF; the 4\textsuperscript{th}--6\textsuperscript{th} and 7\textsuperscript{th}--9\textsuperscript{th} columns denote the intensity and phase of the \vpsf{}, respectively. In \textbf{b} and \textbf{d}, only the fields on the focal plane are shown (see \supp{Supplementary Sections~7.3 and 7.4} for their 3D distribution).
        }
    \label{fig_mixed_state}
\end{figure*}

\medskip\xsubsection{Application in AO-assisted nanoscopy.}
In optical nanoscopy, adaptive optics (AO) allows for better resolution~\cite{Booth2014AdaptiveOpticalMicroscopy} by compensating for the aberrations at the objective's BFP. Our method demonstrates its applicability, particularly in stimulated emission depletion (STED) nanoscopy~\cite{Blom2017StimulatedEmissionDepletion}. The standard AO-assisted STED nanoscopy involves either optimizing the phase component of the VPF represented by Zernike polynomials using hill-climbing method~\cite{Gould2012AdaptiveOpticsEnables, Zdankowski2020NumericallyEnhancedStimulated, Tu2022AccurateBackgroundReduction}, which needs to capture nearly a hundred images, or employing Shack--Hartman wavefront sensor with increased system complexity~\cite{Velasco20213DSuperresolutionDeeptissue}. These methods, which rely on the incomplete scalar model and fail to access the \insitu{} VPF, prove useful to some extent. However, since a high-NA objective is of the essence in STED nanoscopy, further enhancement is limited. The VPF's phase can't be faithfully retrieved without considering its amplitude and polarization~\cite{Hao2010EffectsPolarizationDeexcitation, He2023VectorialAdaptiveOptics}, as well as the imperfections in the detection path. In contrast, as a more complete and accurate approach, ours harnesses AO better.

In our configuration, the depletion \ipsf{} was created by tightly focusing a circularly polarized vortex beam. With the \insitu{} retrieved VPF, we achieved enhanced STED imaging by correcting the phase aberration. In Figs.~\ref{fig_ao_sted}a--c, two adjacent microtubules are hardly distinguishable in confocal microscopy and STED nanoscopy without AO, but are clearly resolved after aberration correction. This improvement is further supported by their line profiles (Fig.~\ref{fig_ao_sted}g). As for the two-color imaging (Figs.~\ref{fig_ao_sted}d--f), the structures of the nuclear pore complex (NPC) and Golgi apparatus are also more clearly resolved with AO. These improvements originate from the more circular and hollow depletion \ipsf{} after aberration correction (Figs.~\ref{fig_ao_sted}i,j), which is further supported by the line profiles in Fig.~\ref{fig_ao_sted}h. More specifically, our method enhances the resolution of the microtubules and two-color imaging by approximate \qty{15}{\percent} and \qty{25}{\percent}, respectively, taking the images without AO for reference. The resolution is evaluated by decorrelation analysis~\cite{Descloux2019ParameterfreeImageResolution}.

\medskip\xsubsection{Mixed-state problems.}
Mixed-state problems are ones for which the \vpsf{} results from the incoherent and/or coherent superposition of multiple beams. Mixed-state problems are always complicated due to increased ambiguity and lower SNR (\supp{Supplementary Sections~7.1--7.2}). While the demixing of scalar waves in low-NA systems is possible~\cite{Thibault2013ReconstructingStateMixtures, Attal2022MixedstateCodedDiffraction}, fully vectorial demixing in high-NA systems remains uncharted. Here we demonstrate the demixing of \vpsf{} containing two states. Notably, we employed 20 pairs of phases for encoding with different distributions between the two states in each pair to alleviate the ambiguity.

We first present experimental results showing the demixing of an incoherent superposition of radially and azimuthally polarized beams (Figs.~\ref{fig_mixed_state}a,b). These results match well independently reconstructed data shown in Fig.~\ref{fig_psf_reconstruction}, reaching around \qty{80}{\percent} accuracy of the predicted two-channel \ipsf{}s without phase encoding. This accuracy is lower than that of single-state reconstruction (\qty{90}{\percent}) since each state suffers from increased ambiguity and lower SNR.

Our method is also effective for coherent cases such as the \vpsf{} created using a 4Pi microscope architecture~\cite{Hell1992Properties4PiConfocal} (Fig.~\ref{fig_mixed_state}c). The coherent utilization of two opposing objectives enhances the axial resolution in microscopy~\cite{Liu2018BreakingAxialDiffraction, Hao2021ThreedimensionalAdaptiveOptical}. The simulated results are shown in Fig.~\ref{fig_mixed_state}d, where both the reconstructed \vpsf{} and VPFs well reflect their ground truth with \qty{80}{\percent} accuracy, including the phase difference between the two VPFs.

We can readily expand our ability of field manipulation in the single-objective architecture to the 4Pi one. For example, we repeated the design of a ring-shaped continuous pattern with a complicated 3D polarization, matching the target shown in Fig.~\ref{fig_psf_manipulation}f. Remarkably, this \vpsf{} generated by 4Pi focusing (\qty{90}{\percent} accuracy; see \supp{Supplementary Fig.~S21}) is more accurate than the result from one single objective (\qty{62}{\percent} accuracy; see \supp{Supplementary Fig.~S25}). The 4Pi architecture offers double degrees of freedom and dramatically extends the bandwidth of the \vpsf{} along the \(z\)-axis, enabling the creation of more intricate \vpsf{}. Similarly, we present another demonstration with a more complicated 3D polarization structure with \qty{88}{\percent} accuracy. Notably, if polarization is not a priority, we can design the intensity distribution alone. For instance, using a single beam, we created an isotropic hollow \ipsf{} that potentially simplifies isoSTED nanoscope~\cite{Hao2021ThreedimensionalAdaptiveOptical}. These results are provided in \supp{Supplementary Sections~7.5--7.7}.

\xsection{Discussion and conclusion}
Our method attains reconstruction accuracy of roughly \qty{90}{\percent} in our configuration and \qty{80}{\percent} even amid substantial background noise. However, achieving higher accuracy remains contingent upon the SNR. To improve the signal level, it is advantageous to employ brighter probes and optimize the optical setup. Additionally, minimizing the refractive index mismatch in focal space can mitigate noise induced by laser reflection.

Our method requires a phase-modulation device for encoding. If incorporating such a device is impossible, mechanical defocus can serve as an alternative. This alternative is simplified but risky, because it reduces the SNR and may fail to eliminate ambiguities.

In our experimental demonstrations of field manipulation and AO-assisted nanoscopy, we solely manipulated or corrected the VPF's phase. Further improvement is achievable by incorporating the VPF's amplitude and/or polarization (\supp{Supplementary Sections~8.1--8.2}). Nevertheless, it requires a more complicated system at the expense of reduced light efficiency~\cite{He2023VectorialAdaptiveOptics}.

Regarding temporal efficiency, our method needs to acquire 10 pairs of encoded \ipsf{}s for accurate reconstruction. In addition, the decoding algorithm takes several extra seconds for iteration. These requirements render our method insufficient for real-time applications. However, if the VPF's polarization is not intricate, appropriate scalar approximations significantly reduce processing time. For example, with just 5 phases for encoding, scalar reconstruction is completed in approximate \qty{0.05}{\second}, roughly 200 times faster than fully vectorial reconstruction (\supp{Supplementary Section~8.3}). Further reduction in measurements and iterations could be achieved by integrating deep learning techniques~\cite{Wang2024UseDeepLearning}. Nevertheless, these alternatives compromise accuracy.

In summary, we have presented a method for \insitu{} fully vectorial tomography and pupil function retrieval of \vpsf{} from 2D encoded \ipsf{}s, as confirmed by both simulations and experiments. By employing elaborately designed phase encoding in the focusing path and splitting polarization in the detection path, our method effectively eliminates ambiguity. The fully vectorial information is then efficiently decoded using analytically derived gradients and matrix implementation of the Fourier transform with minimal samples. Our method is notably robust against aberrations, benefiting from the focus scanning strategy and decoding algorithm. We have demonstrated this method for \insitu{} creating complicated 3D polarization structures and correcting aberrations in optical nanoscopy. Our method also proves applicable for resolving the mixed-state challenges.

Our development provides access to the properties of the electric field, a fundamental representation of light, at the nanoscale in practical deployment. It also represents an important step toward characterizing and optimizing more advanced parameters in experiments, such as energy flow, angular momentum, and optical force.

\bibliographystyle{naturemag}
\footnotesize
\bibliography{\refpath}

\clearpage
\normalsize
\xsection{Methods}
\xsubsection{Phase design for encoding.}
Phase diversity can be achieved by defocus or vanilla random phase, as described in previous studies~\cite{Gonsalves2018PhaseDiversityMath, Wu2019WISHWavefrontImaging, Gutierrez-Cuevas2024VectorialPhaseRetrieval}. However, these methods often encounter challenges with limited diversity and low SNR. For accurate reconstruction, it is essential that the phases used for encoding are distinct enough to minimize ambiguity, while also ensuring that the \ipsf{}s remain within a confined FOV to maintain a high SNR.

Our phase design is informed by Fourier optics, specifically the relationship between the extension of the \ipsf{} and the spatial gradient of the VPF's phase~\cite{Wei2023ModelingOffaxisDiffraction}. However, directly generating a phase with a specific gradient constraint is challenging. Therefore, we propose a solution that involves randomly generating two smooth matrices to represent the initial gradient distribution along the \(x\) and \(y\) directions, constrained within a specified range. It is crucial to account for the inherent phase gradient present in the VPF, which can cause the \ipsf{} to extend beyond the size of an aberration-free Airy disk. To address this, we exclude this intrinsic phase gradient, which is estimated beforehand from the initial \ipsf{} size without phase encoding. Assuming the initial \ipsf{} size is \(l_x \times l_y\), and the target FOV is \(L_x \times L_y\), the phase pattern is designed to expand the \ipsf{} along each axis by \(\Delta x = 0.5 (L_x - l_x) \) and \( \Delta y = 0.5 (L_y - ly)\), respectively. According to Fourier optics, the maximum allowable phase gradient is expressed as
\begin{equation}
    \begin{aligned}
        \abs{\frac{\partial \phi\left(k_x,k_y\right)}{\partial k_{\gamma} }}_{\max} = \Delta \gamma,
    \end{aligned}
    \label{eq_phase_gradient}
\end{equation}
where \(\gamma \in \left\{x,y\right\}\). \(\left(k_{x}, k_{y}\right)\) are the coordinates in the BFP, represented by spatial frequency. With this gradient constraint, the phase for encoding is generated by solving an inverse gradient problem~\cite{Farneback2007EfficientComputationInverse}.

\medskip\xsubsection{Probe sample preparation.}
A gold nanosphere with a diameter of \qty{150}{\nm} was used as the probe. These gold nanospheres (A11-150-CIT-DIH-1-25, Nanopartz) were sparsely deposited onto a coverslip (CG15CH2, Thorlabs), and subsequently embedded in oil (IMMOIL-F30CC, Olympus). The coverslip was then fixed onto a glass slide using nail polish, which was allowed to cure for \qty{24}{\hour}.

\medskip\xsubsection{Biological samples preparation.}
For microtubule labeling, U2OS (human osteosarcoma cell line) cells were purchased from the American Type Culture Collection and cultured in McCoy's 5A medium (Thermo Fisher Scientific) supplemented with \qty{10}{\percent} (v/v) FBS. The cells were maintained at \qty{37}{\degreeCelsius} in a humidified \qty{5}{\percent} CO\textsubscript{2} environment and seeded into glass-bottom dishes (NEST Scientific) at a density of \(1.5\sim\num{2.0d4}\) cells per well before labeling. Following overnight incubation, the cells were washed three times with phosphate-buffered saline (PBS; Thermo Fisher Scientific), fixed with \qty{3}{\percent} (m/v) paraformaldehyde (Electron Microscopy Sciences) and \qty{0.1}{\percent} (v/v) glutaraldehyde (Sigma-Aldrich) for \qty{10}{\minute} at \qty{37}{\degreeCelsius}, and then quenched with NaHB\textsubscript{4} for \qty{7}{\minute} at room temperature (RT). The cells were subsequently incubated with \qty{0.2}{\percent} (v/v) Triton X-100 (Sigma-Aldrich) and \qty{5}{\percent} goat serum (Thermo Fisher Scientific) for \qty{1}{\hour} at RT. Microtubules were stained with mouse anti-\(\alpha\)-tubulin (AF2827; Beyotime Biotechnology) and mouse anti-\(\beta\)-tubulin (AF2835; Beyotime Biotechnology) at \(1 \colon 100\) dilution in PBS overnight at \qty{4}{\degreeCelsius}, and followed by goat anti-mouse STAR RED (STRED-1001-500UG; Abberior GmbH) at \(1\colon 200\) dilution in PBS for \qty{1}{\hour}. Before imaging, the samples were washed three times with PBS for \qty{3}{\minute} each.

NPC and Golgi apparatus samples (Cells 4C NPC-STAR RED; Golgi-STAR ORANGE; Actin-STAR GREEN; DAPI) were purchased from Abberior.

\medskip\xsubsection{Code availability}\newline
The open-source code of the algorithms will be available upon acceptance.

\medskip\xsubsection{Data availability}\newline
The data used to reproduce the results of this study are available from the corresponding authors upon reasonable request.

\medskip\xsubsection{Acknowledgments}\newline
We thank Yu Wang, Dazhao Zhu, Chuankang Li, and Yuran Huang for fruitful discussions, and Xin Luo, Wenwen Gong, Wenli Tao, and Lu Yang for helping prepare samples, and Yunpeng Wang for helping with experiments. This work is funded by National Key R\(\&\)D Program of China (2022YFB3206000), and Scientific and Technological Innovation Project of China Academy of Chinese Medical Sciences (CI2023C009YG).

\medskip\xsubsection{Author contributions}\newline
Xin L. conceived the project, developed the methodology, performed the simulations, and conducted the experiments. S.T. helped with the optical setup. S.T. and Yiwen H. helped with the experiments. Yubing H. helped with sample preparation. X.H. supervised the project. All the authors contributed to the preparation of the manuscript.

\medskip\xsubsection{Competing interests}\newline
The authors declare no competing interests.

\end{document}